\documentstyle[12pt]{article}

\newcommand{\rf}[1]{(\ref{#1})}

\renewcommand{\thefootnote}{\fnsymbol{footnote}}

\newcommand{\newsection}{    
\setcounter{equation}{0}
\section}
\def\appendix#1{
  \addtocounter{section}{1}
  \setcounter{equation}{0}
  \renewcommand{\thesection}{\Alph{section}}
  \section*{Appendix \thesection\protect\indent \parbox[t]{11.715cm} {#1} }
  \addcontentsline{toc}{section}{Appendix \thesection\ \ \ #1}
  }

\def \td {\tilde }
\def \b{\beta}

\def \foot {\footnote}
\def \bi{\bibitem}
\def \la {\label}
\def \tr {{\rm tr}}
\def \ha {{1 \over 2}}
\def \ep{\epsilon}
\def \CC{{\cal C}}
\def \ov {\over}
\def \JJ {{\cal J}}

\def\nline{\,\nabla\kern -0.7em\raise0.2ex\hbox{/}\,\,}
\def\yline{\,y\kern -0.47em /}
\def\aline{\,a\kern -0.49em /}
\def\parline{\,\partial\kern -0.55em /\,\,}

\def\apr{{a^\prime}}
\def\bpr{{b^\prime}}
\def\cpr{{c^\prime}}
\def\dpr{{d^\prime}}
\def\alpr{{\alpha^\prime}}
\def\bepr{{\beta^\prime}}
\def\gapr{{\gamma^\prime}}
\def\depr{{\delta^\prime}}

\def\aha{{\hat{a}}}
\def\bha{{\hat{b}}}
\def\cha{{\hat{c}}}
\def \H {{\cal H}}
\def \t {\theta}
\def \s{\sigma}
\def \d {\partial}

\def\NPB#1(#2)#3{{\it Nucl. Phys.} {\bf B#1} (#2) #3}
\def\PRD#1(#2)#3{{\it Phys. Rev.} {\bf D#1} (#2) #3}
\def\PLB#1(#2)#3{{\it Phys. Lett.} {\bf B#1} (#2) #3}

\def\RMP#1(#2)#3{{\it Rev. Mod. Phys.} {\bf #1} (#2) #3}
\def\MPLA#1(#2)#3{{\it Mod. Phys. Lett.} {\bf A#1} (#2) #3}
\def\CQG#1(#2)#3{{\it Class. Quantum Grav.} {\bf #1} (#2) #3}
\def\AP#1(#2)#3{{\it Ann. Phys.} {\bf #1} (#2) #3}
\def\SJNP#1(#2)#3{{\it Sov. J. Nucl. Phys.} {\bf #1} (#2) #3}

\def \del{\partial}
\def \m {\mu}

\def \four{{\textstyle {1\ov 4}}}
 
\def\det{\hbox{det}}
\def\be{\begin{equation}}
\def\ee{\end{equation}}

\def \ci {\cite}
\def \g {\gamma}
\def \G {\Gamma}

\begin{document}

\begin{titlepage}
\begin{flushright}
FIAN/TD/98-21\\
Imperial/TP/97-98/44  \\
NSF-ITP-98-055\\
hep-th/9805028\\
\end{flushright}
\vspace{.5cm}

\begin{center}
{\LARGE   Type IIB  superstring  action  in
  AdS$_5 \times $S$^5$   background   }\\[.2cm]
\vspace{1.1cm}
{\large R.R. Metsaev${}^{{\rm a,}}$\footnote{\ E-mail: metsaev@lpi.ac.ru}
and A.A. Tseytlin${}^{{\rm a,b,c,}}$\footnote{\ E-mail: tseytlin@ic.ac.uk} }\\
\vspace{18pt}
${}^{{\rm a\ }}${\it
Department of Theoretical Physics, P.N. Lebedev Physical
Institute,\\ Leninsky prospect 53, 117924, Moscow, Russia
}\\
${}^{{\rm b\ }}${\it Blackett Laboratory,
Imperial College, London SW7 2BZ, U.K.}\\
${}^{{\rm c\ }}${\it Institute of Theoretical Physics,
University of California,\\
Santa Barbara, CA 93106, USA}

\end{center}
\vskip 0.6 cm

\begin{abstract}
We construct the  covariant $\kappa$-symmetric  superstring action 
for a type IIB superstring on  $AdS_5\otimes S^5$ 
background.  The action is defined as 
a 2d $\sigma$-model   on the  coset superspace
$SU(2,2|4)\over SO(4,1)\times SO(5)$
 and  is shown to be  the  unique
one that has the correct bosonic and flat space limits.
\end{abstract}

\end{titlepage}
\setcounter{page}{1}
\renewcommand{\thefootnote}{\arabic{footnote}}
\setcounter{footnote}{0}

\newsection{Introduction}
After the construction of   IIB supergravity
 (motivated by the development of superstring theory)
 \cite{S1,S2,W1} it was
immediately realized that in addition to the flat ten-dimensional
space $R^{1,9}$ this theory allows
 $AdS_5\otimes S^5$ (+ self-dual 5-form background) as  another
  maximally supersymmetric `vacuum' \cite{S2}.
  Some  aspects of this  compactification on $S^5$ 
were  studied in \cite{GM1,N1,Ts1}
(this led to the construction of $N=8$ gauged supergravity
in 5 dimensions which describes the `massless' modes \ci{FI}).
In particular,  it was understood \cite{GM1}
that the  Kaluza-Klein modes 
 fall into unitary irreducible representations
of $ N=8, D=5$ anti de Sitter supergroup $SU(2,2|4)$
(which is the same  as the $N=4$  superconformal group in 4
dimensions \ci{HLS}).
The  supergroup  $SU(2,2|4)$ (with the even part
$SU(2,2) \otimes SU(4)\simeq SO(4,2) \otimes SO(6) $ 
which is the isometry of the  $AdS_5\otimes S^5$ space)
thus plays here   the same central role
as  does the  Poincare supergroup  in the flat vacuum.

Motivated by the recent duality conjecture between
the type IIB string theory on $AdS_5\otimes S^5$ background
and $N=4,D=4$ Super Yang-Mills theory  \ci{MA}--\ci{FFZ},  one   would like
to  study  the corresponding string theory directly, using
the world-sheet
methods. This may allow  
 to prove that  $AdS_5\otimes S^5$ space is an exact
string solution, 
 define the  corresponding the 2d conformal theory,
find the   spectrum of  string states,
etc.

Since the $AdS_5\otimes S^5$  space is supported by the  Ramond-Ramond
5-form background,  the NSR approach does not seem 
 to apply in a straightforward way   (while  the
non-local RR vertex operator
is known  in the flat space \ci{FMS}, it is most likely
not sufficient
 to determine the complete form of the NSR
string action when  the space-time metric  is curved).

The manifestly supersymmetric 
Green-Schwarz (GS)  approach \ci{GSS} seems  a more  adequate
one when the RR fields are non-vanishing.
  While the formal superspace
expression for
the  GS superstring action in a generic  type IIB  background
(satisfying the supergravity equations of motion  to guarantee the
$\kappa$-invariance of the string action \ci{WWW})
was  presented in \ci{HOW} (see also \ci{GA}),
it  is not very practical  for finding  the explicit form
of the   action in terms of the coordinate fields $(x,\theta$):
given a particular  bosonic background, one is first to
determine explicitly the corresponding $D=10$ type IIB superfields
which is a complicated problem
 not solved so far in any non-trivial case.\footnote{In fact,
the only known example of a covariant GS action
in a curved RR background was recently constructed in \ci{RT}
(in the case of a non-supersymmetric IIA  magnetic 7-brane background)
using an indirect method based on starting with the known
 supermembrane action \ci{Ber1}
in flat $D=11$ space.}

The remarkable (super)symmetry of the $AdS_5\otimes S^5$
background suggests that here one should apply  an alternative approach which
explicitly uses   the special properties of this vacuum.
Our aim  below will be  to  find   the counterpart of the  covariant
 GS  action in flat space   for the type IIB
string propagating in $AdS_5\otimes S^5$ spacetime
by starting directly with the supergroup $SU(2,2|4)$
and constructing a  $\kappa$-symmetric
2d $\sigma$-model   on the  coset superspace
$SU(2,2|4)\over SO(4,1)\otimes SO(5)$.
The method is  conceptually
  the same
  as used in
\ci{HM}  to reproduce   the flat-space  GS  action
 as  a  WZW-type $\sigma$-model
on the coset superspace  ($D=10$ super Poincare group/Lorentz group $SO(9,1))$.

In Section 2 we shall describe   the structure
of the superalgebra $su(2,2|4)$
and define the invariant Cartan 1-forms $L^A$
on the coset superspace $(x,\t)$.\footnote{For some applications of the formalism
of  the Cartan forms  on coset superspaces
see \cite{N2}--\cite{Ber3}.}

In Section  3  we shall
present  the covariant
 $AdS_5\otimes S^5$ superstring action 
 in the coordinate-free form, i.e.
 in terms of the Cartan 1-forms  $L^A$
on the  superspace.  As in the flat space case \ci{GSS,HM}
it  is given by the sum   the  `kinetic' or `Nambu' term
(2d integral of
the  quadratic term in $L^A$)
plus a Wess-Zumino type term  (3d integral of a closed 3-form  $\H$
on the superspace), with the coefficient of the WZ 
term  fixed uniquely by the requirement of 
$\kappa$-symmetry. 
In the zero-curvature (infinite radius) 
limit   the action reduces to the  standard flat space  GS action
\ci{GSS,HM}.

In Section 4 we shall  find   the explicit 2d form
of the action
by choosing a specific parametrization of the Cartan 1-forms
in terms of
the fermionic coordinates
$\theta$.
   The resulting action
  may be viewed as the  unique maximally-supersymmetric and
 $\kappa$-symmetric extension
 of the bosonic string sigma model  with
 $AdS_5\otimes S^5$ as a target space.
It is given by a `covariantization' of the flat-space GS action 
plus terms containing  higher powers of $\t$.
Though we  explicitly present only the $\t^4$ term, 
 it is very likely
that after an appropriate 
$\kappa$-symmetry  gauge choice 
the { full} action 
will be determined by the $\t^2$ and $\t^4$ terms only.\foot{Such truncation of a  complicated  covariant  GS action 
to $O(\t^4)$ expression was found 
to happen  after choice of the light-cone gauge $\g^+\t=0$ in 
the case of a  curved RR background  considered in \ci{RT}.}

Some 
properties of the  resulting string theory
 will be briefly discussed in Section 5.
The  string action  depends on 
 generalised  `supersymmetric' spinor  covariant derivative
  and  thus   contains  the expected  coupling 
$(\del x \del x \bar \t \g...\g\t\ e^\phi F_{...}$) 
to the RR background.
Since the action is  uniquely determined
by the $SU(2,2|4)$ symmetry, its 
  classical 2d conformal invariance  
  should  survive at the quantum
level: as  in the WZW model case,
  the symmetries of the action prohibit any  deformation
  of its  structure (provided, of course, the regularisation scheme
  preserves these symmetries).
 The central role played by $SU(2,2|4)$
 implies  that not only the  `supergravity' (marginal)
 but also all `massive'  string vertex operators
will  belong to  its
 representations. 

In Appendix A we shall introduce explicit parametrisation of the coset superspace
and  define the corresponding Cartan superconnections. 
In  Appendix B we shall  explain the procedure to compute 
the  expansion of the supervielbeins in powers of  $\t$
which is used to determine higher-order terms in the 
component 2d string action.

\newsection{ $su(2,2|4)$  superalgebra}
Our staring point is the supergroup $SU(2,2|4)$.
Since the string  we are interested in propagates
on the coset
superspace $SU(2,2|4) \over SO(4,1)\otimes SO(5)$
 with the even part being
 $AdS_5\otimes S^5 = {SO(4,2) \over SO(4,1)} \otimes
 {SO(6) \over SO(5)}$
we shall present
the corresponding
superalgebra $su(2,2|4)$
in the
 $so(4,1)\oplus so(5)$  (or  `5+5')  basis.
 The   even generators are then two pairs of
 translations and rotations --
 $(P_a, J_{ab})$ for $AdS_5$ and $(P_\apr,J_{\apr\bpr})$
 for $S^5$  and the odd generators are the two
 $D=10$ Majorana-Weyl spinors
 $Q^{\alpha \alpr}_{ I}$.

\subsection{Notation}
In what follows we use the following  convention for
indices:
\begin{eqnarray*}
a,b,c=0,1,\ldots, 4 &\qquad &so(4,1) \hbox{ vector indices ($AdS_5$ tangent space)}\\
\apr,\bpr,\cpr=5,\ldots, 9 &\qquad & so(5) \hbox{ vector
indices ($S^5$ tangent space) }\\
\aha,\bha,\cha =0,1,\ldots, 9& & \hbox{ combination  of }
(a,\apr), (b,\bpr), (c,\cpr) \hbox{\ ($D=10$  vector indices)}
\\
\alpha,\beta,\gamma,\delta =1,\ldots,4 &\qquad &
so(4,1) \hbox{ spinor indices ($AdS_5$)}\\
\alpr,\bepr,\gapr,\depr =1,\ldots,4 &\qquad
&so(5) \hbox{ spinor indices ($S^5$) }\\
\hat \alpha,\hat \beta,\hat \gamma  =1,\ldots, 32& & \hbox{$D=10$  MW spinor indices}
\\
I,J,K,L=1,2 & & \hbox{ labels  of the two sets of  spinors}
\end{eqnarray*}
Similarly, $\hat \mu= (\mu, \mu')$  will denote the coordinate indices of 
$AdS_5\otimes S^5$.
The generators of the $so(4,1)$ and $so(5)$
Clifford algebras  are $4\times 4$ matrices  $\gamma_a$  and  $\gamma_\apr$
$$
\gamma^{(a} \gamma^{b)}=\eta^{ab}=(-++++)\,,
\qquad
\gamma^{(\apr} \gamma^{\bpr)}=\eta^{a'b'}=(+++++)\ .
$$
It will be useful to define    also the    ten
$4\times 4$ matrices $\hat \gamma^{\hat a}$
\be \hat{\gamma}^a\equiv \gamma^a\,,
\qquad
\hat{\gamma}^\apr\equiv {\rm i}\gamma^\apr \ .
\ee
 We shall assume that
$
(\gamma^a)^\dagger=\gamma^0 \gamma^a \gamma^0\,,
\ \
(\gamma^\apr)^\dagger=\gamma^\apr
$
and  that the
Majorana condition  is
diagonal with respect to the  two supercharges
\begin{equation}\label{mcon}
\bar Q_{\alpha\alpr I} \equiv (Q^{\beta \beta'}_{  I})^\dagger
(\gamma^0)_\alpha^\beta \delta^{\beta'}_{\alpha'}
=-Q^{\beta\beta'}_{ I}C_{\beta\alpha}C_{\beta'\alpr} \ .
\end{equation}
Here  $C=(C_{\alpha\beta})$ and $C^\prime=(C_{\alpha'\beta'})$
  are  the
charge conjugation matrices\foot{To simplify the notation,
we shall put primes
on matrices and generators to distinguish between
the objects corresponding to the
two factors ($AdS_5$ and $S^5$)   only in the cases
when they do not carry explicit (primed) indices.} of the
 $so(4,1)$ and $so(5)$
Clifford algebras  which are used to raise and lower spinor
indices, e.g.,  $
Q_{\alpha\alpr I}\equiv
Q^{\beta\bepr}_{ I}C_{\beta\alpha}C_{\bepr\alpr}.
$
The bosonic generators will be assumed to be   antihermitean:
$
P_{a}^\dagger=- P_a\,,\  P_\apr^\dagger=- P_\apr \,,
\
J_{ab}^\dagger=- J_{ab}\,,\ J_{\apr\bpr}^\dagger=- J_{\apr\bpr}.
$
We shall use
the   2$\times $2 matrices
$\epsilon^{{IJ}}= - \epsilon^{{JI}}, \ \epsilon^{12}=1,$ 
and $s^{IJ}\equiv {\rm diag} (1,-1)$
to contract the indices $I=1,2$.\footnote{As in the flat space
case \ci{GSS,HM},
$s^{IJ}$  will appear in the WZ term of the GS action, indicating the
breakdown of the formal $U(1)$ symmetry  between the two
Majorana-Weyl   spinors of the same chirality,
which a symmetry of the type  IIB superfield supergravity \ci{W1} but is 
broken in perturbative string theory, e.g.,  is absent 
in the  action of the  superstring 
 in  a type IIB supergravity background
 \ci{HOW}.
}
Unless otherwise stated, we shall always assume the summation rule
 over the repeated indices (irrespective of their
position).

The 10-dimensional 
$32\times 32$ Dirac matrices $\G^{\hat a}$  of  $SO(9,1)$
($\G^{(\hat a}\G^{\hat b)}=\eta^{\hat a \hat b}$) 
and the corresponding charge conjugation matrix $\CC$
can be  represented as
\be
\G^a=\gamma^a\otimes I\otimes \sigma_1 \,, \qquad
\G^\apr=I\otimes\gamma^\apr\otimes \sigma_2\,,
\qquad  \CC =C\otimes C^\prime \otimes {\rm i}\sigma_2\,,
     \ee
where $I$ is the $4\times 4$ unit matrix and $\sigma_i$ are the Pauli
matrices.
Note that $C\gamma^{a_1\ldots a_n}$ are symmetric (antisymmetrc)
for $n=2,3$ mod 4 ($n=0,1$ mod 4). The same properties are valid
for $C^\prime\gamma^{a_1^\prime\ldots a_n^\prime}$.

A $D=10$ positive chirality  32-component  spinor $ \Psi$ is   decomposed
   as follows:
$
\Psi =\psi\otimes \psi^\prime \otimes  \left({1\atop 0}\right)
$  (16-component spinors $\theta^I$ and $L^I$ below will correspond to 
32-component spinors of positive chirality). 
The  Majorana condition $\bar \Psi \equiv \Psi^\dagger \G^0=
\Psi^T \CC$  then takes the same form
as in (\ref{mcon}) but with sign plus (for a negative chirality spinor
$\sim \left({0\atop 1}\right)$  the Majorana condition 
is the same as in \rf{mcon}). 
A useful formula
which explains   the 10-dimensional origin of some of the   expressions
below  is
\be
K^\aha \bar{\Psi}_1 \Gamma^\aha \Psi_2
=K^a \bar{\chi}_1\gamma^a \chi_2  +{\rm i} K^\apr \bar{\chi}_1
\gamma^\apr \chi_2\,,
\ee
where $\Psi_n$ $(n=1,2$)
are the $D=10$ Majorana-Weyl    spinors of  positive  chirality,
$\chi_n=\psi_n\otimes \psi'_n$, 
and $K^\aha$ is a 10-vector.
Here (and in  similar expressions below) 
$\gamma^a$ and $\gamma^\apr$ stand for $\gamma^a \otimes I$ and 
$I\otimes \gamma^\apr$.

\subsection{Commutation relations }
 The
 commutation relations  for the generators
 $ T_A =(P_a,P_\apr,J_{ab},J_{\apr\bpr},
 Q_{\alpha\alpha'I})$ are\foot{Because of
the  presence of $\epsilon^{IJ}$ some of the anticommutators
are not diagonal with
respect to the supercharges.  This is  related to the
standard  choice of diagonal  Majorana condition  (2).
 By
obvious redefinitions we could make the commutation relations
diagonal with respect two the supercharges but this would lead to
a
non-diagonal Majorana condition.}
$$
[P_a,P_b]=J_{ab}\,,
\qquad\qquad\qquad
[P_\apr,P_\bpr]=-J_{\apr\bpr}\,,
$$
$$
[P_a,J_{bc}]=\eta_{ab}P_c-\eta_{ac}P_b\,,
\qquad
[P_\apr,J_{\bpr\cpr}]=\eta_{\apr\bpr}P_\cpr -\eta_{\apr\cpr}P_\bpr\,,
$$
$$
[J_{ab},J_{cd}]=\eta_{bc}J_{ad}+3\hbox{ terms}\,,
\qquad
[J_{\apr\bpr},J_{\cpr\dpr}]
=\eta_{\bpr\cpr}J_{\apr\dpr}+3\hbox{ terms}\,,
$$
$$
[Q_I,P_a]=-\frac{{\rm i}}{2}\epsilon_{{IJ}}
Q_J\gamma_a \,,
\qquad
[Q_I,P_\apr]
=\frac{1}{2}\epsilon_{{IJ}} Q_J\gamma_\apr\,,
$$
\be
[Q_I,J_{ab}]=-\frac{1}{2} Q_I\gamma_{ab}\,,
\qquad
[Q_I,J_{\apr\bpr}]=-\frac{1}{2}Q_I\gamma_{\apr\bpr} \,,
\ee
\begin{eqnarray*}
\{Q_{\alpha \alpr I }, Q_{\beta \bepr J}\}
&=&\delta_{{IJ}}
\bigg[-2{\rm i}C_{\alpr\bepr}(C\gamma^a)_{\alpha\beta} P_a
+2C_{\alpha\beta}(C^\prime\gamma^\apr)_{\alpr\bepr}
P_\apr\bigg]
\\
&+&\epsilon_{{IJ}}
\bigg[C_{\alpr\bepr}(C\gamma^{ab})_{\alpha\beta} J_{ab}
-C_{\alpha\beta}(C^\prime\gamma^{\apr\bpr})_{\alpr\bepr}
J_{\apr\bpr}\bigg] \ .
\end{eqnarray*}
The curvature (radius $R$) parameter of
$AdS_5 \otimes S^5$ space  can be introduced by rescaling the generators
$
P_a \to { R}P_a, \ P_\apr \to R P_\apr, \
 J \to J,    \   Q_I \to \sqrt{R} Q_I  .
$
Then the limit $R\to \infty$  gives the subalgebra of
$D=10$ type IIB Poincare
superalgebra.\foot{The Poincare superalgebra contains
the generators $J_{a\apr}$ which are absent in $su(2,2|4)$:
the $SO(2,4) \otimes SO(6)$ isometry of
$AdS_5 \otimes S^5$ leads 
in the limit $R\to \infty$ only to a
subgroup of $SO(9,1)$ (times translations).
While  the $AdS_5 \otimes S^5$ background  has maximal  number 32
of Killing spinors, it  does not have maximal number  55  of Killing vectors
in 10 dimensions (the dimension of $SO(2,4) \otimes SO(6)$ is 30).}

\subsection{Cartan  1-forms   }
To find the
the super-invariant and
 $\kappa$-invariant  string action we will use  the formalism of
Cartan forms defined on the coset superspace.
The left-invariant Cartan  1-forms
$$L^A = dX^M L^A_{M}\ ,  \qquad \ \ \
X^M=(x, \theta)$$
are given by
\begin{equation}\label{expan1}
G^{-1}dG = L^A  T_A
\equiv L^aP_a+L^\apr P_\apr+\ha L^{ab}J_{ab}+\ha L^{\apr\bpr}J_{\apr\bpr}
+L^{\alpha \alpr I}Q_{\alpha \alpr I} \ ,
\end{equation}
where $G= G({x,\theta})$ is a coset representative in $SU(2,2|4)$.
 A specific choice of
$G(x,\theta)$ which we shall use in Section 4 is described in Appendix A.

$L^a$ and $L^\apr$ are the 5-beins,  $L^{\alpha\alpr I}$ are the two
spinor 
 16-beins
and  $L^{ab}$ and $L^{\apr\bpr}$ are the
Cartan connections. They satisfy
the  Maurer-Cartan equations implied by the
structure of the  $su(2,2|4)$ superalgebra
\begin{eqnarray}
\label{mcf}
&&
dL^a=-L^b\wedge L^{ba}
-{\rm i}\bar{L}^{^I}\gamma^a\wedge L^{^I}
\,, \qquad \ \ \
dL^\apr=-L^\bpr\wedge L^{\bpr\apr}
+\bar{L}^{^I}\gamma^\apr\wedge L^{^I}\,,
\\
&&
dL^{ab}=-L^a\wedge L^b-L^{ac}\wedge L^{cb}
+\epsilon^{^{IJ}} \bar{L}^{^I}\gamma^{ab}\wedge
L^{^J}\,,
\\
\label{mcl}
&&
dL^{\apr\bpr}=L^\apr\wedge L^\bpr
-L^{\apr\cpr}\wedge L^{\cpr\bpr}
-\epsilon^{^{IJ}}\bar{L}^{^I}\gamma^{\apr\bpr}\wedge
L^{^J} \ ,
\end{eqnarray}
\be
dL^{^I}
=-\frac{{\rm i}}{2}\gamma^a\epsilon^{^{IJ}}L^{^J}\wedge L^a
+\frac{1}{2}\epsilon^{^{IJ}}\gamma^\apr L^{^J}\wedge L^\apr
+\frac{1}{4}\gamma^{ab}L^{^I}\wedge L^{ab}
+\frac{1}{4}\gamma^{\apr\bpr}L^{^I}\wedge L^{\apr\bpr} \ .
\ee
The rescaling of the generators  which restores
the scale   parameter $R$ of $AdS_5 \otimes S^5$
corresponds to
$L^a \to { R^{-1}}L^a, \ L^\apr \to R^{-1} L^\apr, \
 L^{ab} \to L^{ab},   \  L^{a'b'} \to L^{a'b'}
 , \ \   L^I \to { R^{-1/2}} L^I.$

For comparison,
 let us note that  in  the flat superspace case
\be
G({x,\t}) = {\rm exp} ( x^{\hat a} P_{\hat a}  + \t^I Q_I)\,,
\qquad [P_{\hat a}, P_{\hat b}]=0\,,
\ \ \  \{ Q_I, Q_J\} = -2{\rm i}  \delta_{IJ} (\CC\Gamma^{\hat a}) P_{\hat a}\,,
\la{supf}
\ee
and  thus  the coset space
vielbeins  in $G^{-1} d G = L^A T_A$ are given by
\be
L_0^{\hat a} = d x^{\hat a}  - i \bar \t^I \G^{\hat a}d\t^I \,, \qquad
L_0^I = d \t^I \ .
\la{flat}
\ee

\newsection{Superstring action as  $SU(2,2|4) \over
SO(4,1) \otimes SO(5)$  superspace  sigma model }

Our aim below  will be  to  construct the superstring action
that satisfies the following conditions
(some of which are not completely independent):

(a) its bosonic part is the standard $\sigma$-model
with the $AdS_5 \otimes S^5$ as a  target space;

(b) it  has global $SU(2,2|4)$ super-invariance;

(c) it is invariant under  local $\kappa$-symmetry;

(d)  it reduces to the  standard   Green-Schwarz type IIB
superstring action

\ \ \  \ \ in the flat-space
($R\to \infty$) limit.

\noindent
We  shall  find   that  such action exists and  is
{\it unique}.   Its
   leading $\t^2$ fermionic term
will  contain
the  required  coupling to the RR 5-form field background.
This is, of course, expected as the
$\kappa$-symmetry implies the satisfaction of the
IIB supergravity equations of motion \ci{HOW}
but the  unique  supergravity solution  with the metric of
$AdS_5 \otimes S^5$   has  a  non-trivial  $F_5$
background
($F_5 \sim \epsilon_5$ in each of the two factors).

It is useful to recall that the flat-space
GS superstring  action may be written
in the  manifestly supersymmetric form
in terms of  the 1-forms   (\ref{flat})
as a sum of the `kinetic' term and a WZ term (integral of a
closed  3-form) \ci{HM}\foot{This action  may be compared with the standard bosonic
WZW action on a group
manifold: in a similar notation, if
 $ G^{-1} dG = L^A T_A$, \ $\tr (T_A T_B) = c_{AB}, \ \tr (T_A [T_B, T_C]) =
 f_{ABC}$,
 then \ci{WW}
 $$I_{\rm WZW} = k\bigg[
 \int_{\del {M_3}}  d^2\s  \sqrt{g} g^{ij} c_{AB} L^A_i L^B_j
 +  
{ 1 \over 6} {\rm i} 
 \int_{M_3}  f_{ABC} L^A \wedge L^B \wedge L^C\bigg]\ . $$
}
\be
I_0=-\frac{1}{2}\int_{\del {M_3}}  d^2\sigma\  \sqrt{g}\ g^{ij}\
 L_{0i}^{\hat a}  L_{0j}^{\hat a}
+  {\rm i}\int_{M_3}
 s^{IJ}   L^{\hat a}_0 \wedge  \bar{L}_0^I\Gamma^{\hat a}
\wedge  L^J_0
\ ,
\la{actif}
\ee
where   $s^{IJ}$   is   defined by
$s^{11}=-s^{22}=1,\ s^{12}=s^{21}=0$
and the string tension is set to be 
 ${1\over 2\pi \alpha'}=1$.
$g_{ij}$ ($i,j=0,1$) is a world-sheet metric with signature
$(-+)$\ ($g=- \det g_{ij}$). 
The coefficient of the WZ term is fixed by the condition of local $\kappa$-invariance \ci{GSS}.
Using (\ref{flat}) one observes  that the  3-form in the WZ term is exact and
thus finds  the  explicit 2d form of the GS action \ci{GSS}
$$ I_0=\int d^2 \sigma\ {\cal L}_0 = 
  \int d^2 \sigma\  \bigg[ - \ha \sqrt{g} g^{ij} (\del_i x^{\hat a} -
{\rm i} \bar \t^I \G^{\hat a} \del_i \t^I)
(\del_j x^{\hat a} -
{\rm i} \bar \t^J \G^{\hat a} \del_j \t^J)
$$ \be
 -\   {\rm i} \ep^{ij} s^{IJ}   \bar \t^I \G^{\hat a} \del_j  \t^J
(
  \del_i x^{\hat a} -  \ha {\rm i}   \bar \t^K \G^{\hat a}  \del_i \t^K)
 \bigg] \ ,
 \la{GRE}
\ee
in which the   $\ep^{ij}$-term is invariant under global supersymmetry
only up to a total derivative.
The action we shall find below is
the generalisation of (\ref{GRE})  to the case  when
the  free bosonic $\del^i x^\aha \del_i x^\aha $ term
is replaced by the $\s$-model on the $AdS_5 \otimes S^5$ space.

\subsection{General structure  of the action }
As the flat-space action,   the  action for the
 type IIB superstring
propagating in $AdS_5 \otimes S^5$ space-time
will be  given by a sum of the `$\sigma$-model' term $I_{\rm kin}$
and a WZ term $I_{\rm WZ}$ which is the integral of a closed 3-form
$\H$ over a 3-space  $M_3$ which has the world-sheet as its boundary,
\be
I= I_{\rm kin} +  I_{\rm WZ} \,,
\qquad \ \
I_{\rm WZ}={\rm i}\int_{M_3}  \H \,, \qquad \ \    d \H =0 \ . \la{gene}
\ee
To satisfy the condition  of
$SU(2,2|4)$ invariance both $I_{\rm kin}$
and $\H$   should be constructed
in terms of the Cartan 1-forms $L^A$.
The basic  observation
is that 
under the action of an arbitrary element
of the  isometry group
the vielbeins  transform as tangent
vectors of the stability subgroup. 
 Thus any  invariant of the stability
subgroup ($SO(4,1)\otimes SO(5)$ in the present case)
  constructed in terms of $L^A$
will be  automatically     invariant  under  the full
 isometry group
 ($SU(2,2)|4)$, i.e.
$SO(4,2)\otimes SO(6)$ and supersymmetry).

The structure  of $I_{\rm kin}$ is then  fixed unambiguously
by the conditions (a) and (b) 
\begin{equation}
I_{\rm kin}=-\frac{1}{2}\int d^2\sigma\  \sqrt{g}\ g^{ij}
\ L_i^{\hat a}  L_j^{\hat a} \ .
\end{equation}
Here  
the repeated indices are contracted 
 with  $\eta_{\hat a\hat b}=
(\eta_{ab},\eta_{a'b'}) $
and $L^A_i =  \d_i X^M L^A_{M}$ are the induced components
of the supervielbein.

The only relevant  3-forms  built out of $L^A$
which are invariant  under
$SO(4,1)\otimes SO(5)$   are given by the
following   linear
combination ($k,k' $  are  free parameters)
\begin{equation}\label{tinv}
\H^I  =\ k\  L^a\wedge\bar{L}^I \gamma^a \wedge L^I + \ k'\
L^\apr\wedge\bar{L}^I \gamma^\apr \wedge L^I \ ,\ \ \    \quad I=1,2 \ .
\end{equation}
Here  we
  do {\it not}  sum  over the repeated  indices $I$.
 Using the Maurer-Cartan equations, one finds
 (no summation over $I$)
$$
d(L^a\wedge \bar{L}^I \gamma^a \wedge L^I)
=\epsilon^{IJ}L^a\wedge L^\apr\wedge \bar{L}^I\gamma^a\gamma^\apr\wedge  L^J
-{\rm i}\bar{L}^I\gamma^a\wedge  L^I\wedge  \bar{L}^J\gamma^a\wedge  L^J 
$$
\be
-\ {\rm i}\epsilon^{IJ}L^a\wedge L^b\wedge  \bar{L}^I\gamma^{ab} \wedge L^J
\,,
\label{d1}
\ee
$$
d(L^\apr\wedge \bar{L}^I \gamma^\apr\wedge  L^I)
={\rm i}\epsilon^{IJ}L^a\wedge L^\apr\wedge \bar{L}^I\gamma^a\gamma^\apr\wedge  L^J
+\bar{L}^I\gamma^\apr\wedge  L^I\wedge  \bar{L}^J\gamma^\apr\wedge  L^J 
$$
\be
+\ 
\epsilon^{IJ}L^\apr \wedge L^\bpr \wedge \bar{L}^I\gamma^{\apr\bpr}\wedge L^J
\la{d2}\,,
\ee
so   that to cancel  the  terms in $d\H^I$
which are given by  the first terms in  the r.h.s of
 (\ref{d1}),(\ref{d2})
we are to  put $ k' ={\rm i} k$. Then
\begin{equation}\label{3form}
\H^I =\  k \ (  L^a\wedge  \bar{L}^I\gamma^a\wedge  L^I
+{\rm i} L^\apr\wedge  \bar{L}^I\gamma^\apr\wedge  L^I)  \ , \ \ \ \  \
 \ \   I=1,2\
\end{equation}
$$
d\H^I=-{\rm i} k\ ( \bar{L}^I\gamma^a \wedge L^I\wedge\bar{L}^J\gamma^a\wedge  L^J
- \bar{L}^I\gamma^\apr \wedge L^I\wedge\bar{L}^J\gamma^\apr \wedge L^J)  
$$
\be
-\ {\rm i}k\ \epsilon^{IJ}(L^a\wedge L^b \wedge \bar{L}^I\gamma^{ab} L^J
-L^\apr\wedge L^\bpr \wedge \bar{L}^I\gamma^{\apr\bpr} \wedge L^J)\ . 
\la{d3}
\ee
It is
 easily verified that the only possibility  to obtain a
closed  3-form is to consider
\be
\H\equiv \H^1-\H^2 \ . \la{forma}
\ee
To prove this one
uses the  Fierz identity
$
(C\gamma^a)_{\alpha\delta}(C\gamma^a)_{\gamma\beta}=
2(C_{\alpha\beta}C_{\gamma\delta}
-C_{\alpha\gamma}C_{\beta\delta})
$
for $SO(4,1)$ and   
$\bar{L}^1\gamma^{ab}\wedge  L^2 = -\bar{L}^2 \gamma^{ab}\wedge  L^1
$
 and  similar relations   for $SO(5)$  part.\foot{Let us note that the first line in  the expression (\ref{d3}) can be
rewritten in terms of ten-dimensional spinors and
$\Gamma$-matrices  only,   and, as in the flat-space case \ci{GSS,HM},
  the  fact that
$d\H=0$ is a consequence
 of the  famous identity for the $D=10$
Dirac matrices
$
\G^\aha_{\hat \alpha(\hat{\beta}}
\G^\aha_{\hat{\gamma}\hat{\sigma})}=0$ (first line in \rf{d3})
and the relation
$\bar{L}^1\gamma^{ab}\wedge L^2 = -\bar{L}^2 \gamma^{ab} \wedge L^1$ (second line in \rf{d3}).}

In the flat-space limit $\H$ reduces to the
3-form  in the GS action (\ref{actif}).
As in flat space, the value of the
 overall coefficient  $k$ in $\H$  is fixed to be 1
by the requirement of $\kappa$-symmetry of the whole action
 (which is proved in the  next subsection).
The final expression for  the action
written
 in the  $SU(2,2|4)$ invariant
form  in terms of the vielbeins $L^a,L^{a'}$ and $L^{I}$
 thus has the same structure as    the GS action (\ref{actif})
\be
I=-\frac{1}{2}\int_{\del {M_3}}  d^2\sigma\  \sqrt{g}\ g^{ij}
 L_i^\aha L_j^\aha
+  {\rm i}\int_{M_3}
 s^{IJ}  L^\aha \wedge  \bar{L}^I\hat \gamma^\aha \wedge  L^J
\ ,
\la{actio}
\ee
or, explicitly,
\be
I=-\frac{1}{2}\int_{\del {M_3}}  d^2\sigma\  \sqrt{g}\ g^{ij}
( L_i^a L_j^a + L_i^\apr L_j^\apr)
+  {\rm i}\int_{M_3}
 s^{IJ} (  L^a\wedge  \bar{L}^I\gamma^a\wedge  L^J
+{\rm i} L^\apr\wedge  \bar{L}^I\gamma^\apr\wedge  L^J)
\ ,
\la{acti}
\ee
and, indeed, reduces to    (\ref{actif}) in the flat-space limit.

Since the   3-form  $\H$  is closed,  in a local coordinate system 
it
can be  represented
   as $\H= d {\cal B}$; then
  the action takes the  usual   2d $\s$-model form which will
  be  considered in Section 4.

\subsection{Invariance under $\kappa$-symmetry and equations of motion}
The action (\ref{acti})
 is invariant with respect to
the  local  $\kappa$-transformations \ci{SS,GSS}
which is useful to write down in terms of
  $
\delta x^a\equiv \delta X^M L_{M}^a\,,
\  \delta x^\apr\equiv \delta X^M L_{M}^\apr\,, \
\delta \theta^I\equiv \delta X^M L_{M}^I$
\begin{eqnarray}\label{ktrans}
&&
\delta_\kappa x^a =0\,,
\qquad \delta_\kappa x^\apr =0\,,
\qquad \ \ \delta_\kappa \theta^I=
2(L_i^a \gamma^a -  {\rm i}L_i^\apr\gamma^\apr)\kappa^{iI} \ ,
\\
&&
\delta_\kappa(\sqrt{g}g^{ij})
=-16{\rm i}\sqrt{g}(
P_-^{jk}\bar{L}_k^1\kappa^{i1}
+P_+^{jk}\bar{L}_k^2\kappa^{i2}) \ .
\end{eqnarray}
Here
$
P_\pm^{ij} \equiv\frac{1}{2}(g^{ij}\pm
\frac{1}{\sqrt{g}} \epsilon^{ij}) \ ,
$
and 16-component spinor $\kappa^{iI}$ (the corresponding  32-component
 spinor  has opposite chirality to that of $\theta$)  
  satisfy the  (anti) self duality
constraints
\be
P_-^{ij} \kappa_j^1=\kappa^{i1}\,,
\qquad
P_+^{ij} \kappa_j^2=\kappa^{i2}\,, 
\ee
which can be rewritten as
$
\frac{1}{\sqrt{g}}\epsilon^{ij}\kappa_j^1=-\kappa^{i1},
\
\frac{1}{\sqrt{g}}\epsilon^{ij}\kappa_j^2=\kappa^{i2}.
$
To demonstrate $\kappa$-invariance   one uses
the following
expressions for the variations of the  Cartan 1-forms:
\begin{eqnarray*}
\delta L^a
&=&d\delta x^a+L^{ab}\delta x^b+L^b\delta x^{ba}
+2{\rm i}\bar{L}^I\gamma^a\delta \theta^I\,,
\\
\delta L^\apr
&=&d\delta x^\apr+L^{\apr\bpr}\delta
x^\bpr+L^\bpr\delta x^{\bpr\apr}
-2\bar{L}^I\gamma^\apr\delta \theta^I\,,
\\
\delta L^I&=&d \delta \theta^I
+\frac{{\rm i}}{2}\epsilon^{IJ}(\delta x^a\gamma^a
+{\rm i}\delta x^\apr\gamma^\apr)L^J
-\frac{{\rm i}}{2}\epsilon^{IJ}(L^a\gamma^a
+{\rm i}L^\apr\gamma^\apr)\delta\theta^J
\\
&-&\frac{1}{4}(\delta x^{ab} \gamma^{ab}
+\delta x^{\apr\bpr}\gamma^{\apr\bpr})L^I
+\frac{1}{4}(L^{ab}\gamma^{ab}+L^{\apr\bpr}\gamma^{\apr\bpr})
\delta\theta^I\,,
\end{eqnarray*}
where
$
\delta x^{ab}\equiv \delta X^M L_{M}^{ab}\,,
\ \delta x^{\apr\bpr}\equiv \delta X^M L_{M}^{\apr\bpr}.
$
The crucial  relation  that allows one to check the
 $\kappa$-invariance
of the action (and also to obtain  the equations of motion)
directly in terms of  the coordinate-invariant  Cartan forms is
\begin{equation}\label{extra}
\delta \H=d \Lambda \,, \qquad \ \ \  \Lambda\equiv\Lambda^1-\Lambda^2 \ ,
\end{equation}
\begin{equation}\label{g2}
\Lambda^I\equiv  \bar{L}^I \gamma^a \wedge L^I\delta x^a
+2L^a\wedge \bar{L}^I\gamma^a\delta \theta^I
+{\rm i} \bar{L}^I \gamma^\apr \wedge L^I\delta x^\apr
+2{\rm i}L^\apr\wedge \bar{L}^I\gamma^\apr\delta \theta^I\,.
\end{equation}
The equations of motion  that follow from the  action (\ref{acti})
 are
\begin{eqnarray}\label{eqqs}
&&
\sqrt{g}g^{ij}(\nabla_i L_j^a+L_i^{ab}L_j^b)
+{\rm i}\epsilon^{ij}s^{IJ}\bar{L}_i^I\gamma^a L_j^J
=0\,,
\\
&&
\sqrt{g}g^{ij}(\nabla_i L_j^\apr+L_i^{\apr\bpr}L_j^\bpr)
-\epsilon^{ij}s^{IJ}\bar{L}_i^I\gamma^\apr L_j^J
=0\,, \\
&&
(\gamma^a L_i^a+{\rm i}\gamma^\apr L_i^\apr) (\sqrt{g} g^{ij} \delta^{IJ} - \epsilon^{ij}
s^{IJ})       L_j^J=0\,,
\end{eqnarray}
where $\nabla_i$ is the  $g_{ij}$-covariant derivative.
These relations should be supplemented  by   the standard constraint
\be L^a_i L^a_j+L^\apr_i L^\apr_j = \ha g_{ij} g^{kl}
 (L^a_k L^a_l+L^\apr_k L^\apr_l) \ ,
\ee
following from the variation of the action
 over $g_{ij}$.

Note that like the `2d+3d' form (\ref{acti})
 of the action (but not its   2d  form discussed in the next Section, cf. (\ref{GRE})),
the equations of motion
  admit a manifestly
  covariant representation in terms
  of the   Cartan 1-forms.
    There  is a certain  similarity
  with the  equations of motion   of  the WZW
  model.
In   the conformal gauge $\sqrt{g} g^{ij}=\eta^{ij}$
these equations (like in the case of the
$\sigma$-model on $G/H$ space)
imply the existence of  dim $G$ conserved
currents.  
The `chirality' (presence of $\ep^{ij}$-terms)
of
the above equations  has  purely fermionic nature.
More direct analogy with the WZW model may be possible after certain
bosonisation of the fermionic degrees of freedom.

\newsection{ Explicit 2-dimensional  form of the  action}
To find the  explicit   form of the action 
in terms  of the coordinate 2d field $\t$ 
 which generalises (\ref{GRE})  we  are  to 
 choose a 
particular  parametrization of the coset representative $G$ in
(\ref{expan1}):
\be 
G(x,\theta) = g(x) {\rm g} (\t)\ , \ \ \ \ \ \ \ \ \
{\rm g} (\t) ={\rm exp}
 (\t^I Q_I)
\ . \la{prr}
\ee
 Here 
$g(x)$ is a coset representative of
$[SO(4,2)\otimes SO(6)]/[SO(4,1)\otimes SO(5)]$, i.e.  $x=(x^\mu,x^{\mu'})$ provides a 
certain parametrization of $AdS_5\otimes S^5$  which  we 
will not  need  to  specify  in what follows.

To
represent the WZ term in \rf{gene},\rf{actio} as an integral over
the 2-dimensional space we  use the standard  trick
of rescaling
$\theta\rightarrow \t_t\equiv t\theta$, \
\be
I_{\rm WZ} =  I_{\rm WZ}(t=1) \,, \qquad \ \
I_{\rm WZ}(t)
={\rm i}\int_{M_3}   \H_t\,, \ \ \ \ \ \  \H_t = \H (\t_t) \ .
\ee
Then  (\ref{extra})  implies
\be
\partial_t I_{\rm WZ}(t)
={\rm i} \int_{\del M_3} \ \partial_t \Lambda \ ,
\qquad \ \
\partial_t\Lambda  =-2s^{IJ}L_t^\aha
\bar{\theta}^I\hat{\gamma}^\aha L_t^J\ ,
\ee
where $L^A_t \equiv L^A(\t_t)$.  We have used (\ref{g2})
and that
$\partial_t
\theta_t=\theta$ and $\partial_t x^\aha=0$.
Thus
\begin{equation}\label{nsys}
I_{\rm WZ}=-2{\rm i}
\int_0^1 dt\int d^2\sigma\ \epsilon^{ij} s^{IJ}  L_{it}^\aha
\bar{\theta}^I\hat{\gamma}^\aha L_{jt}^J \ .
\end{equation}
Eq. (\ref{nsys}) together with  equations from
Appendices A and B provide a setup for a systematic
calculation of the action  $I$ \rf{acti} as an expansion in powers of $\t$.

The expansions of the  Cartan 1-forms are given by
(see Appendix B, cf. \rf{flat})
\begin{eqnarray}\la{expanf}
L^a
&&=  e^a
-{\rm i}\bar{\theta}^I\gamma^a D\theta^I
+ \frac{1}{12}{\rm i}\epsilon^{IJ}(\bar{\theta}^I\theta^J
\bar{\theta}^K\gamma^a D \theta^K
-\bar{\theta}^I\gamma^a\gamma^\apr\theta^J
\bar{\theta}^K\gamma^\apr D \theta^K)
\nonumber \\
&& + \  \frac{ 1}{24}
{\rm i} \epsilon^{KL}(-\bar{\theta}^I\gamma^{abc}\theta^I
\bar{\theta}^K\gamma^{bc}D \theta^L
+\bar{\theta}^I\gamma^a\gamma^{\bpr\cpr}\theta^I
\bar{\theta}^K\gamma^{\bpr\cpr} D \theta^L)+ ...\,,
\\
L^\apr
&&= e^\apr
+\bar{\theta}^I\gamma^\apr D\theta^I   + \frac{1}{12}\epsilon^{IJ}(\bar{\theta}^I\theta^J
\bar{\theta}^K\gamma^\apr D\theta^K
-\bar{\theta}^I\gamma^a\gamma^\apr\theta^J
\bar{\theta}^K\gamma^a D \theta^K)
\nonumber\\
&& + \  \frac{1}{24}
\epsilon^{KL}(-\bar{\theta}^I\gamma^{\apr\bpr\cpr}\theta^I
\bar{\theta}^K\gamma^{\bpr\cpr} D \theta^L
+\bar{\theta}^I\gamma^\apr\gamma^{bc}\theta^I
\bar{\theta}^K\gamma^{bc} D \theta^L)+ ... \,,
\\
\la{expanl}
L^I
&&=D\theta^I + \frac{1}{6}\epsilon^{IJ}(-\gamma^a \theta^J
\bar{\theta}^K\gamma^a D\theta^K
+\gamma^\apr\theta^J \bar{\theta}^K\gamma^\apr D\theta^K)
\nonumber\\
&&
+\ \frac{1}{12}\epsilon^{KL}(\gamma^{ab}
\theta^I \bar{\theta}^K\gamma^{ab} D\theta^L
- \gamma^{\apr\bpr}\theta^I
\bar{\theta}^K\gamma^{\apr\bpr} D\theta^L)+ ...\,,
\end{eqnarray}
where $e^a$, $e^\apr$, $\omega^{ab}$, $\omega^{\apr\bpr}$
are 5-beins and Lorentz connections of $AdS_5$ and $S^5$,
and  the generalised spinor  covariant differential  $D\t^I$
is defined by\foot{The 
remarkable $D^2=0$ property (see Appendix A)
is the condition of integrability of  the Killing spinor equation
$D_{\mu} \epsilon^I=0$.}
\be
D\theta^I\equiv D^{IJ}\theta^J\ , \ \ \ \ \ \ \  \ D^{IJ}D^{JK}=0\,,
\ee
\be
D^{IJ}\equiv  \delta^{IJ} {\cal D} - \frac{1}{2} {\rm i} \epsilon^{IJ}e^\aha\hat \gamma^\aha =
 \delta^{IJ}\bigg[d
+\frac{1}{4}(\omega^{ab}\gamma^{ab}
+\omega^{\apr\bpr}\gamma^{\apr\bpr})\bigg]
- \frac{1}{2} {\rm i} \epsilon^{IJ}(e^a\gamma^a
+ {\rm i}  e^\apr\gamma^\apr) \ .
\la{dee}
\ee
The  $\t^2$ and $\t^4$
terms in $I_{\rm WZ}$ are determined by
\be
\partial_t^2\Lambda|_{t=0}=-2e^\aha \wedge s^{IJ}
\bar{\theta}^I\hat{\gamma}^\aha D\theta ^J
\,, \ee
\begin{eqnarray*}
\partial_t^4\Lambda|_{t=0}
&&= -24{\rm i} \bar{\theta}^1\hat{\gamma}^\aha D\theta ^1\wedge
\bar{\theta}^2\hat{\gamma}^\aha D\theta ^2
-2e^\aha\wedge s^{IJ} \bar{\theta}^I\hat{\gamma}^\aha
\partial_t^3L_t^J|_{t=0}
\\
&&=-24{\rm i} \bar{\theta}^1\hat{\gamma}^\aha D\theta ^1\wedge
\bar{\theta}^2\hat{\gamma}^\aha D\theta ^2
+4e^a\wedge\bar{\theta}^1\gamma^{ab}\theta^2\bar{\theta}^I
\gamma^bD\theta^I
-4{\rm i}
e^\apr\wedge\bar{\theta}^1\gamma^{\apr\bpr}\theta^2\bar{\theta}^I
\gamma^\bpr D\theta^I
\\
&& + \ 
s^{IJ}\epsilon^{KL}
e^a\wedge(-\bar{\theta}^I\gamma^{abc}\theta^J
\bar{\theta}^K \gamma^{bc} D\theta^L
+\bar{\theta}^I\gamma^a\gamma^{\bpr\cpr}\theta^J
\bar{\theta}^K \gamma^{\bpr\cpr} D\theta^L)
\\
&& + \ 
{\rm i}s^{IJ}\epsilon^{KL}
e^\apr\wedge(\bar{\theta}^I\gamma^{\apr\bpr\cpr}\theta^J
\bar{\theta}^K \gamma^{\bpr\cpr} D\theta^L
-\bar{\theta}^I\gamma^\apr\gamma^{bc}\theta^J
\bar{\theta}^K \gamma^{bc} D\theta^L) \ .
\end{eqnarray*}
Using  these expressions  we find the following
result for the  action \rf{acti}
\be
 I = \int d^2\sigma \   {\cal L} \ ,
\ \ \ \ \ \ \   \   {\cal L}={\cal L}_{1}+{\cal L}_{2}+ O(\t^6)
\,,
\ee
where ${\cal L}_{1}$ is essentially a `covariantisation' of
the flat-space
 GS  Lagrangian, i.e. it
 has the same structure as  ${\cal L}_{0}$ in  \rf{GRE}
 but with $\del_i \t \to D_i \t$
and 5-beins contracting  target-space indices
$$
{\cal L}_{1}
=-\frac{1}{2}\sqrt {g} g^{ij}
(e_i^\aha  -{\rm i}\bar{\theta}^I\hat{\gamma}^\aha D_i\theta^I)
(e_j^\aha  -{\rm i}\bar{\theta}^I\hat{\gamma}^\aha D_j\theta^I)
$$
\be
-\ {\rm i}\epsilon^{ij} e_i^\aha( \bar{\theta}^1 \hat{\gamma}^\aha
D_j\theta^1  - \bar{\theta}^2\hat{\gamma}^\aha
D_j\theta^2)
+\epsilon^{ij}\bar{\theta}^1\hat{\gamma}^\aha D_i\theta ^1
\bar{\theta}^2\hat{\gamma}^\aha D_j\theta ^2
\,,
\la{stan}
\ee
where
\be
e_i^\aha = (e^a_i, e^\apr_i)\,,\qquad
  \ e^a_i  \equiv e_{\mu }^a (x) \partial_i x^{\mu}\,, 
\qquad \ e^\apr_i  \equiv e_{\mu' }^\apr (x') \partial_i x^{\mu'}
\ .\ee
The   additional   $\t^4$ term ${\cal L}_{2}$
is  given by
\be
{\cal L}_{2} = {\cal L}_{2{\rm kin}} + {\cal L}_{2{\rm WZ}} \,,
\ee
\begin{eqnarray*}
{\cal L}_{2{\rm kin}}
&&=
 { 1 \over 24} \sqrt {g} g^{ij}\bigg({\rm i} e^a_i \bigg[2\epsilon^{IJ}
(-\bar{\theta}^I\theta^J
\bar{\theta}^K\gamma^a D_j \theta^K
+\bar{\theta}^I\gamma^a\gamma^\apr\theta^J
\bar{\theta}^K\gamma^\apr D_j \theta^K)\\
&&
+\
\epsilon^{KL}(\bar{\theta}^I\gamma^{abc}\theta^I
\bar{\theta}^K\gamma^{bc}D_j \theta^L
-\bar{\theta}^I\gamma^a\gamma^{\bpr\cpr}\theta^I
\bar{\theta}^K\gamma^{\bpr\cpr} D_j \theta^L)\bigg]
\\
&&
+\  e^\apr_i \bigg[ 2\epsilon^{IJ}  (-\bar{\theta}^I\theta^J
\bar{\theta}^K\gamma^\apr D_j\theta^K
+\bar{\theta}^I\gamma^a\gamma^\apr\theta^J
\bar{\theta}^K\gamma^a D_j \theta^K)
\\&&
+\  \epsilon^{KL}(\bar{\theta}^I\gamma^{\apr\bpr\cpr}\theta^I
\bar{\theta}^K\gamma^{\bpr\cpr} D_j \theta^L
-\bar{\theta}^I\gamma^\apr\gamma^{bc}\theta^I
\bar{\theta}^K\gamma^{bc} D_j \theta^L)\bigg] \bigg)\,,
\end{eqnarray*}
\begin{eqnarray*}
{\cal L}_{2\rm WZ} &&=
\frac{1}{24}\epsilon^{ij} \bigg[\ 4
({\rm i}e_i^a\bar{\theta}^1\gamma^{ab}\theta^2\bar{\theta}^I
\gamma^bD_j\theta^I
+e_i^\apr\bar{\theta}^1\gamma^{\apr\bpr}\theta^2\bar{\theta}^I
\gamma^\bpr D_j\theta^I)
\\
&&+\ 
 {\rm i}s^{IJ}\epsilon^{KL}
e_i^a(-\bar{\theta}^I\gamma^{abc}\theta^J
\bar{\theta}^K \gamma^{bc} D_j\theta^L
+\bar{\theta}^I\gamma^a\gamma^{\bpr\cpr}\theta^J
\bar{\theta}^K \gamma^{\bpr\cpr} D_j\theta^L)
\\
&&+\ 
 s^{IJ}\epsilon^{KL}
e_i^\apr(-\bar{\theta}^I\gamma^{\apr\bpr\cpr}\theta^J
\bar{\theta}^K \gamma^{\bpr\cpr} D_j\theta^L
+\bar{\theta}^I\gamma^\apr\gamma^{bc}\theta^J
\bar{\theta}^K \gamma^{bc} D_j\theta^L) \bigg] \ .
\end{eqnarray*}
We have not determined explicitly 
 the $\t^6$ terms in $I$ (this is, in principle, 
 straightforward
using the method explained above), but it is likely
that after an appropriate
$\kappa$-symmetry  gauge choice the expression
for  the action will contain only $\t^4$ terms at most, i.e.
the {\it full} action
will be given  by ${\cal L}_{1}+{\cal L}_{2}$
presented   above.

\newsection{Some  properties of the action}

The presence of the  `mass'  term  linear in $\g^\aha$ 
in the covariant derivative 
$D$ \rf{dee}  may be interpreted as being due  to a
non-trivial self-dual 5-form background
and is directly connected with the 
global supersymmetry of the action.
Essentially the same  derivative   $D_{\hat\mu}=\del_{\hat\mu} + \four
 \omega^{\hat a\hat b}_{\hat \m}
\Gamma_{\hat a\hat b}  + c \G^{\hat \m_1...\hat
\m_5} \G_{\hat \m}\  e^\phi F_{\hat \m_1...\hat \m_5}$   appears in the 
 Killing spinor equation of type IIB supergravity
(see, e.g., \ci{KKK}).

The  leading-order 
coupling to the RR background field 
 $\del x^{\hat \lambda} \del x^{\hat \nu}
 \bar \t \Gamma_{\hat \lambda} \Gamma^{\hat \mu_1 ... \hat \mu_5}
 \Gamma_{\hat \nu} 
\t\ e^\phi F_{\hat \mu_1 ... \hat \mu_5}$
is, indeed, contained in the 
 $e^\aha \bar \t\g^\aha  D \t$  term in \rf{stan}\foot{Note that  
$\gamma^{a_1\ldots a_5}={\rm i} \epsilon^{a_1\ldots a_5}\,,
\ 
\gamma^{a'_1\ldots a'_5}= \epsilon^{a'_1\ldots a'_5}, 
$
and $(F_5)_{AdS_5} = Q \epsilon_5$, 
 $(F_5)_{S^5} = Q \epsilon_5$, where $Q$ is related to the RR charge.
Also, $e^\phi=g_s=$const and the radius $R \sim  (g_s Q)^{1/4}$
is set equal to 1.}
$$
{\cal L}_{1}(\t^2) =
{\rm i} (\sqrt{g} g^{ij} \delta^{IJ} - \epsilon^{ij}
s^{IJ})  e^\aha_i \bar{\theta}^I\gamma^\aha D_i\theta^J
$$
\be 
\ \to \ \ ... + 
\frac{1}{2}
(\sqrt{g} g^{ij} \delta^{IK} - \epsilon^{ij}
s^{IK}) 
  \epsilon^{KJ} e^\aha_{\hat \mu}
 e^\bha_{\hat \nu} \del_i x^{\hat \mu} \del_j x^{\hat \nu}
  \bar \t^I\g^\aha  \gamma^\bha \t^J 
\ . \la{RRR}
\ee 
 The presence of  terms of 
higher orders in $\t$ reflect  the curved  nature
of the  background metric. 

The uniqueness of the action  which is completely 
fixed by the requirement of  $SU(2,2|4)$ 
supersymmetry  suggests,  by analogy with the  WZW model, 
 that  it defines an exact 2d  conformal  field
theory in which  non-conformal-invariance 
of the bosonic coset space 
$AdS_5 \otimes S^5$ $\sigma$-model 
is compensated by contributions from the fermionic terms.
 The fact  that the action contains 
 no free parameters
(except for the radius of the background)
that can be renormalised  implies  that the corresponding 
$AdS_5 \otimes S^5$ + 5-form
background  should be an {\it exact}  string solution.
 In general,  the  global symmetry implies that only the 
coefficients in front of the `kinetic'  and WZ terms
in \rf{actio}  may be renormalised. 
However, the analogy with WZW model suggests that 
global (super)symmetry should prohibit 
renormalisation of the coefficient in front of the
 {\it WZ term}  (which is manifestly supersymmetric 
 when written
in the global 3d form).\foot{This is related to the  fact that
 the coefficients in the 3-form $\H$ in \rf{3form},\rf{forma}
 are covariantly constant
(recall that the contributions to the $\beta$-function
of  the $B_{mn}$-coupling in the 
standard bosonic 
sigma model always involve derivatives of $H=dB$).}
But then the local  $\kappa$-symmetry relating the two terms in 
\rf{actio} should imply
  that the coefficient of the  `kinetic'
term is also not renormalised,\foot{In the WZW model case  similar
fact is 
effectively  related
(via   the  Polyakov-Wiegmann identity)  to  the
 existence of on-shell
conserved chiral currents and associated  affine symmetry \ci{WW,deal}.}
 i.e. 
that \rf{actio} is a conformal model.
The remaining question  is why its central charge  is 
not shifted from  its flat-space value.
Possible corrections to the dilaton $\beta$-function
are proportional to   invariants constructed from  curvature 
and $F_5$  and are constant in  the  given background.
The crucial point is that the leading-order correction
vanishes since $R= (F_5)^2 =0$ because of the 
 cancellation between the $AdS_5$ and $S^5$ 
contributions.  Higher-order corrections should vanish 
in a special `$\kappa$-symmetric' scheme.\foot{The situation should be similar to that in the case of  the  NS-NS  $AdS_3 \otimes S^3$ 
background described by 
supersymmetric  $SL(2,R) \times SU(2)$ WZW model (with both factors 
having the same level $k$).
In this  case \ci{CT} the non-trivial part of the central charge is 
$c= { 3 k_+ \over k_+-2} + {3  k_- \over k_- +2}= 
(3 + { 2 \over k }) + ( 3 - { 2 \over k })=6  $,
where the leading-order correction cancels between 
the two factors and the shifts of the levels $k_+=k+2, \  k_-=k-2$
(implying the absence of all higher-order corrections)
take place  in a special supersymmetric scheme.}

The absence of both perturbative and non-perturbative
 \ci{BG} string
corrections to this vacuum (see also \ci{KR}) may be related to  the 
32 fermionic zero modes  associated with the global supersymmetry
 of the action (implying, as in flat space,  the existence of
certain non-renormalisation theorems).

One may  check the conformal invariance 
directly by a   one-loop calculation 
(using a `quantum' $\kappa$-symmetry  `light-cone' gauge 
$\Gamma^+ \theta=0$ 
as in   \ci{gris}).  It is fairly clear that   the 
RR coupling  \rf{RRR} will produce an extra 
$F^2_5$ contribution to the renormalisation of the 
$(\del x)^2$ term in the action, leading to the required  
 conformal invariance condition (supergravity equation of motion) $R_{\hat\m\hat\nu} \sim $ $e^\phi (F^2_5)_{\hat\m\hat\nu}$.

Since the  action  we constructed  has 
manifest $SU(2,2|4)$ 
symmetry, the  string spectrum  
should be classified by  representations of 
this supergroup.  In particular, the  marginal 
vertex operators should be in 
one-to-one correspondence  with  the unitary irreducible 
representations  of $SU(2,2|4)$
associated  \cite{GM1} with the modes of type IIB supergravity 
on $AdS_5 \otimes S^5$.

Some  basic  features of this relation can be seen 
directly from the   main  ${\cal L}_{1}$ \rf{stan} 
part of the action.  Here we make only  qualitative  comments
to indicate  some implications  
  of the presence of the RR background on the 
conditions on  vertex operators.   
 Let us formally ignore the curvature of the background 
 and fix  the l.c. gauge:
$x^+\sim \tau, \ \Gamma^+ \theta =0$. 
One can then  show that 
the presence of the  above  RR vertex \rf{RRR}
induces the mass term in the equation of
conformal invariance for  the graviton 
perturbation $
(\Box + 2)h_{\mu\nu}=0
$
 (this is directly related to the computation of the 1-loop $\beta$-function mentioned above).
The coupling term for  the RR 2-form perturbation  $\td H$ 
together with the 5-form background term \rf{RRR}
contained in the action 
are represented  by\ 
$
 \partial x^\mu \partial x^\lambda
\bar \theta \gamma^\mu \gamma^{\rho\sigma} \theta
\td H_{\rho\sigma\lambda}
+  \partial x^\mu \partial x ^\nu\bar  \theta
\gamma^\mu\gamma^\nu\theta
$, \ or, in the l.c. gauge, by  \ $
\bar \theta \gamma^- \gamma^{ij}\theta \td H_{jk+}
+  \partial x^i \bar \theta \gamma^i \gamma^-\theta$
\ ($i,j,k=1,2,3$).
The resulting additional contribution 
 to the NS-NS 2-form $\b$-function is proportional
to \ \ 
$
\td H_{jk+}\partial x^i <(\bar\theta \gamma^- \gamma^i\theta)
(\bar\theta \gamma^-\gamma^{jk}\theta)>.
$
Since 
 tr$(\gamma^i\gamma^j\gamma^k)=\epsilon^{ijk}$
and $<(\theta\theta)(\theta\theta)>\ \sim \ln \varepsilon$
we get 
$\partial_\mu H_{\mu i +}
=\epsilon_{ijk}\td H_{jk+}$, which is the l.c. gauge
($ H_{\mu\nu-}=0, \ 
\td H_{\mu\nu-}=0$)
version
of the well-known equation \ci{S2}\ 
$\partial_\mu H_{\mu\nu\rho}
\propto \epsilon_{\nu\rho\lambda_1\lambda_2\lambda_3}
\td H_{\lambda_1\lambda_2\lambda_3}$\ 
`mixing' the two 2-form tensors in the presence of the
 5-form background.

This  consequence of the non-vanishing   RR background is
also  directly 
related to the presence  of only one set  of $SO(6)$  gauge 
vectors  among the marginal perturbations. 
This is, of course, consistent with the supergravity spectrum 
of KK compactification on $S^5$ but is different  from 
what one would naively expect on the basis of analogy with the WZW model.\foot{A.T.
is grateful to M. Douglas for raising this issue.}
The reason lies in the chirality of the string action \rf{acti},\rf{stan}
 in the fermionic sector.
The latter implies 
 the existence of only one marginal $SO(6)$ vertex operator
associated with the super-extension of conserved currents of $S^5$ $\sigma$-model: 
in contrast to the group space case,  the 
coset space $S^5$ has only  one copy of $SO(6)$ as an isometry,
with the marginality of the corresponding vertex depending 
on the  non-zero  RR 5-form background.\foot{Put differently, the 
2d parity inversion
 does not lead to a new marginal operator
as in the presence of  the  RR background it cannot be  
accompaneed by a field transformation like $G \to G^{-1}$ 
in the WZW model case.}

Further progress in unraveling  the properties 
of this string  theory  obviously   depends on understanding how to  
 fix  a  suitable  $\kappa$-symmetry gauge which will 
lead to a substanial  simplification of the action.
To be able to define the corresponding 
2d conformal field theory 
one may  need  also  to identify 
the proper variables in which the  world-sheet description
may become more transparent.
These  are expected to  be related to the conserved
  (super)currents. 
In the case of a bosonic $\sigma$-model defined on a coset
 space (e.g., $AdS_5$)  the conserved currents 
$\JJ_i^p= k_a^p e_i^a,
\ 
\JJ_i^{pq}=k_a^{pq} e_i^a
$
are constructed in terms of the  vielbeins (projected to the world sheet)
$e^a_i$ and  the 
 Killing vectors 
$k_\mu^p$,  $k_\mu^{pq}=-k_\mu^{qp}$ 
\ ($\nabla_{(\mu} k_{\nu)}^p=0,
\ 
\nabla_{(\mu} k_{\nu)}^{pq}=0$) \ 
 which generate 
translations and rotations (see in this connection   \cite{BRFR,FGib}).
The  
 currents  are conserved 
but not chiral in general.
We expect that the corresponding 
supercurrents constructed using also Killing spinors
should  play  an  important  role in 
the 2d theory defined by \rf{acti}.

In conclusion, let us mention
that an
approach similar
to the one  described  in this paper can be used to 
construct  actions
of  the  IIB superstring
 on $AdS_3 \times S^3 \times T^4$ space
  or  of  the $D=11$ 
membrane   on $AdS_4 \times S^7$.
In the former case there are two possible choices for a WZ
term   leading to the (1/2 supersymmetric) 
actions corresponding 
to the  cases of non-trivial NS-NS or RR 2-form field
backgrounds.

\bigskip

\setcounter{section}{0}
\setcounter{subsection}{0}
\begin{center}
{\bf Acknowledgments}
\end{center}
A.A.T.  is grateful  to   M. Douglas, D. Gross,
R. Kallosh, H. Ooguri  and A. Polyakov
 for  useful  and stimulating
   discussions.
This  work was supported in part
by PPARC,   the European
Commission TMR programme grant ERBFMRX-CT96-0045,
the NSF grant PHY94-07194, the INTAS grant No.96-538 and
 the Russian Foundation for Basic Research Grant No.96-02-17314.

\appendix{Coset superspace parametrisation   and some 
general relations}
In the  parametrisation $G(x,\t)= g(x) {\rm g}(\t)$ 
\rf{prr} of an element   of 
$SU(2,2|4)/[SO(4,1)\otimes SO(5)]$ 
\ $g(x)$ represents the bosonic
 coset 
$[SO(4,2)\otimes SO(6)]/[SO(4,1)\otimes SO(5)]$
and defines 
\be
 g^{-1}dg =e^\aha P_\aha  
+\frac{1}{2}\omega^{\aha\bha}J_{\aha\bha}
 \  
\ee
the  5-beins $e^\aha=(e^a,e^\apr)$ and Lorentz connections 
$\omega^{   \aha\bha}=(\omega^{ab},\omega^{\apr\bpr})$
of $AdS_5 \otimes S^5$
(here $ J_{\aha\bha}\equiv  (J_{ab}, J_{\apr\bpr})$, 
i.e. $\omega^{\aha\bha}$ and $\Omega^{\aha\bha}$ 
below do not contain   `cross-terms'). 
They satisfy the standard (zero torsion, constant curvature) relations
\be\label{lorcon}
de^a+\omega^{ab}\wedge e^b=0\,, \qquad 
de^\apr+\omega^{\apr\bpr}\wedge e^\bpr=0\,,
\ee
\be
d\omega^{ab}+\omega^{ac}\wedge \omega^{cb}=-e^a\wedge e^b
\,, \qquad 
d\omega^{\apr\bpr}
+\omega^{\apr\cpr}\wedge \omega^{\cpr\bpr}
=e^\apr\wedge e^\bpr \ . 
\ee
On the odd part  of the  superspace it is useful to introduce the
following Cartan forms
\be 
{\rm g}^{-1} d{\rm g}= 
\Omega^\aha P_\aha   
+\frac{1}{2}\Omega^{\aha\bha}J_{\aha\bha}
+\Omega^I Q_I\ , \ \ \ \ \ \ \ \ \ {\rm g}={\rm g}(\t) \ ,  
\ee
where, by definition,  
\begin{equation}\label{decom1}
\Omega^\aha= d\theta^I\Omega_I^\aha\,,
\qquad
\Omega^{\aha\bha}= d\theta^I\Omega_I^{\aha\bha}\,,
\qquad
\Omega^I= d\theta^J\Omega_J^I \ . 
\end{equation}
For the  exponential parametrization $ {\rm g}
= $exp$(\t^IQ_I)$ the Cartan
superconnections satisfy the following 
constraints
\be 
\theta^I \Omega_I^\aha=0\,,
\qquad
\theta^I \Omega_I^{\aha\bha}=0\,,
\qquad
\theta^J \Omega_J^I=\theta^I
\ . \ee
It  is easy  to prove that   
\begin{equation}\label{rep2}
G^{-1}d G  =e^{-\theta Q}D e^{\theta Q} \ , 
\ \ \ \ \ \ \
G(x,\theta) = g(x) {\rm g} (\t)= g(x) e^{\theta Q} 
 \ , 
\end{equation}
where  the covariant differential is given by
\be 
D=d+e^\aha P_\aha+\frac{1}{2}\omega^{\aha\bha}J_{\aha\bha}
\,, \qquad \ \ 
D^2=0
 \ . \la{diff}\ee
Eq. (\ref{rep2}) can be  re-written as follows
\begin{equation}\label{expan2}
G^{-1}d G 
=(e^\aha+\Omega_{cov}^\aha)P_\aha
+\frac{1}{2}(\omega^{\aha\bha}+\Omega_{cov}^{\aha\bha})J_{\aha\bha}
+\Omega_{cov}^IQ_I\,,
\end{equation}
where $\Omega^A_{cov}$ are obtained  from $\Omega^A$ 
by  the replacement  $d\to D$:
\begin{equation}\label{decom2}
\Omega_{cov}^\aha= D\theta^I\Omega_I^\aha\,,
\qquad
\Omega_{cov}^{\aha\bha}= D\theta^I\Omega_I^{\aha\bha}\,,
\qquad
\Omega_{cov}^I= D\theta^J\Omega_J^I\ . 
\end{equation}
The explicit form of the covariant differential  $
D\theta^I\equiv D^{IJ}\theta^J$
\be
D^{IJ}\equiv \delta^{IJ} (d
+\frac{1}{4}\omega^{\aha\bha}\gamma^{\aha\bha})
- \frac{1}{2}{\rm i} \epsilon^{IJ} e^\aha\hat\gamma^\aha
\ ,   \ \ \ \ \ \ \ 
 D^{IJ}D^{JK}=0 \ , 
\la{deed}
\ee
      was already given  in  \rf{dee}. 
Comparing (\ref{expan2}) with  the defining relation (\ref{expan1}), 
we get the following representation  for  the 
Cartan  1-forms in terms
of the 5-beins,  Lorentz connections and superconnections
\begin{equation}\label{carcon}
L^\aha=e^\aha+\Omega_{cov}^\aha\,,
\qquad
L^{\aha\bha}= \omega^{\aha\bha}
+\Omega_{cov}^{\aha\bha}\,,
\qquad
L^I=\Omega_{cov}^I \ . 
\end{equation}
To compute the Cartan 1-forms we are thus
to  calculate  the Cartan superconnections
(\ref{decom1}), make the  covariantization (\ref{decom2}) and then
use the expressions (\ref{carcon}).

\appendix{$\t$-expansion of superconnections and \ \ \  1-forms}

Making the rescaling $\theta\rightarrow t\theta$  we get 
the defining equation
\begin{equation}\label{defeq}
e^{-t\theta Q}  d e^{t\theta Q}
=\Omega_t^\aha P_\aha+\ha \Omega_t^{\aha\bha} J_{\aha\bha}
+\Omega_t^I Q_I\ . 
\end{equation}
The forms $\Omega^A$  we  are interested in  are given by $\Omega^A_t|_{t=1}$.
Eq. \rf{defeq}  implies  the following differential
equations
\be
\label{derf}
\partial_t \Omega_t^a =-2{\rm i}\bar{\theta}^I\gamma^a \Omega_t^I
\,, \qquad 
\partial_t \Omega_t^\apr =2\bar{\theta}^I\gamma^\apr \Omega_t^I
\,, \ee
\be
\partial_t \Omega_t^{ab} =2\epsilon^{IJ}\bar{\theta}^I\gamma^{ab}
\Omega_t^J
\,, \qquad 
\partial_t \Omega_t^{\apr\bpr}
=-2\epsilon^{IJ}\bar{\theta}^I\gamma^{\apr\bpr} \Omega_t^J
\,, \ee
\be
\label{derl}
\partial_t \Omega_t^I
=d\theta^I
-\frac{{\rm i}}{2}\epsilon^{IJ}\gamma^a \theta^J \Omega_t^a
+\frac{1}{2}\epsilon^{IJ}\gamma^\apr \theta^J
\Omega_t^\apr +\frac{1}{4}\gamma^{ab} \theta^I \Omega_t^{ab}
+\frac{1}{4}\gamma^{\apr\bpr} \theta^I \Omega_t^{\apr\bpr}\,, 
\ee
with the initial conditions
\begin{equation}\label{incon}
\Omega_{t}^\aha|_{t=0}=\Omega_{t}^{\aha\bha}|_{t=0}=
\Omega_{t}^I|_{t=0}=0\ . 
\end{equation}
  According to (\ref{defeq}),
 the  power of $\theta$ in the expansion of $\Omega^A$ 
coincides with that of $t$ in the expansion of $\Omega^A_t$.
Making use of  \rf{derf},\rf{derl}
 and (\ref{incon})
we find 
\be\la{nnn}
\partial_t^2 \Omega_t^a|_{t=0}
=-2{\rm i}\bar{\theta}^I\gamma^a d\theta^I
\,, \qquad 
\partial_t^2 \Omega_t^\apr|_{t=0}
=2\bar{\theta}^I\gamma^\apr d\theta^I
\,, \ee \be 
\partial_t^2 \Omega_t^{ab}|_{t=0}
=2\epsilon^{IJ}\bar{\theta}^I\gamma^{ab} d\theta^J
\,, \qquad 
\partial_t^2 \Omega_t^{\apr\bpr}|_{t=0}
=-2\epsilon^{IJ}\bar{\theta}^I\gamma^{\apr\bpr} d\theta^J
\,, \ee
\begin{eqnarray*}
\partial_t^3 \Omega_t^I|_{t=0}
&&=\epsilon^{IJ}(-\gamma^a \theta^J \bar{\theta}^K\gamma^ad\theta^K
+\gamma^\apr\theta^J \bar{\theta}^K\gamma^\apr d\theta^K)
\\
&&+\ \frac{1}{2}\epsilon^{KL}(\gamma^{ab}
\theta^I \bar{\theta}^K\gamma^{ab} d\theta^L
-\gamma^{\apr\bpr}\theta^I
\bar{\theta}^K\gamma^{\apr\bpr} d\theta^L)\ ,
\end{eqnarray*}
\begin{eqnarray*}
\partial_t^4 \Omega_t^a|_{t=0}
&&=2{\rm i}\epsilon^{IJ}(\bar{\theta}^I\theta^J
\bar{\theta}^K\gamma^a d \theta^K
-\bar{\theta}^I\gamma^a\gamma^\apr\theta^J
\bar{\theta}^K\gamma^\apr d \theta^K)
\\
&&+\ {\rm i}\epsilon^{KL}(-\bar{\theta}^I\gamma^{abc}\theta^I
\bar{\theta}^K\gamma^{bc}d \theta^L
+\bar{\theta}^I\gamma^a\gamma^{\bpr\cpr}\theta^I
\bar{\theta}^K\gamma^{\bpr\cpr}d \theta^L)\,,
\\
\partial_t^4 \Omega_t^\apr|_{t=0}
&&=2\epsilon^{IJ}(\bar{\theta}^I\theta^J
\bar{\theta}^K\gamma^\apr d \theta^K
-\bar{\theta}^I\gamma^a\gamma^\apr\theta^J
\bar{\theta}^K\gamma^a d \theta^K)
\\
&&+\ \epsilon^{KL}(-\bar{\theta}^I\gamma^{\apr\bpr\cpr}\theta^I
\bar{\theta}^K\gamma^{\bpr\cpr}d \theta^L
+\bar{\theta}^I\gamma^\apr\gamma^{bc}\theta^I
\bar{\theta}^K\gamma^{bc}d \theta^L) \ . 
\end{eqnarray*}
From  (\ref{carcon}) and \rf{nnn}
we  find  leading terms in the  $\theta$-expansion of the 
  Cartan 1-forms 
\be
L^a= e^a-{\rm i}\bar{\theta}^I\gamma^a D\theta^I+
\ldots
\,, \qquad 
L^\apr= e^\apr
+\bar{\theta}^I\gamma^\apr D\theta^I+\ldots
\,, \ee
\be 
L^{ab}=\omega^{ab}
+\epsilon^{IJ}\bar{\theta}^I\gamma^{ab}D\theta^J
+\ldots\,, \qquad 
L^{\apr\bpr}=\omega^{\apr\bpr}
- \epsilon^{IJ}\bar{\theta}^I\gamma^{\apr\bpr}D\theta^J
+\ldots\,, \ee
\be 
L^I=D\theta^I+\ldots \ . 
\ee
Eq.  (\ref{carcon}) and  the above 
expressions for $\partial_t^n \Omega_t^A|_{t=0}$
allow   one to find
the  higher
order corrections to  the Cartan 1-forms
given in \rf{expanf}--\rf{expanl}.
 These are used to
calculate higher-order terms in  the string  action.



\begin{thebibliography}{30}

\parskip=0.pt





\bibitem{S1}
M.B. Green and J.H. Schwarz,
Phys. Lett. 122B (1983) 143;\\
J.H. Schwarz and P. West,
Phys. Lett. 126B (1983) 301.

\bibitem{S2}
J.H. Schwarz,
Nucl. Phys. B226 (1983) 269.

\bibitem{W1}
P.S. Howe and P.C. West, Nucl. Phys. B238 (1984) 181.

\bibitem{GM1}
M. Gunaydin and N. Marcus,
Class. Quantum Grav. 2 (1985) L11.

\bibitem{N1}
N.J. Kim, L.J. Romans and P. van Nieuwenhuizen,
Phys. Rev. D32 (1985) 389.

\bibitem{Ts1}
T.T. Tsikas,
Class. Quantum Grav. 2 (1985) 733.


\bi{FI}
 M. Gunaydin, L.J. Romans and N.P.
   Warner, 
      Phys. Lett. B154 (1985) 268;\\
   M. Pernici, K. Pilch and  P.
   van Nieuwenhuizen,  Nucl. Phys. B259 (1985)
    460.


\bi{HLS}
R. Haag, J.T. Lopuszanski and M. Sohnius,
Nucl. Phys. B88 (1975) 257.

\bi{MA}
J. Maldacena,
``The Large N Limit of Superconformal Field Theories
and Supergravity",
hep-th/9711200.


\bi{GKP}
S.S. Gubser, I.R. Klebanov and A.M. Polyakov,
``Gauge Theory Correlators
 from Non-Criticial String Theory'',
hep-th/9802109.


\bi{HO}
G. Horowitz and H. Ooguri, 
 ``Spectrum of Large N Gauge Theory from Supergravity",
hep-th/9802116.

\bi{W}
E. Witten,
``Anti de Sitter Space and Holography",
hep-th/9802150.

\bi{FFZ}
 S. Ferrara,   C. Fronsdal and A. Zaffaroni, 
``On N=8 Supergravity on $AdS_5$ and $N=4$
Superconformal Yang-Mills theory", 
hep-th/9802203.


\bi{FMS}
D. Friedan, E. Martinec and S. Shenker, 
Nucl. Phys. B271 (1986) 93.  


\bi{GSS}
M.B. Green and J.H. Schwarz,
Phys. Lett.  B136 (1984) 367;
Nucl. Phys. B243 (1984) 285. 

\bi{WWW}
E. Witten, Nucl. Phys. B266 (1986) 245. 

\bi{HOW}
M.T. Grisaru, P.S. Howe, L. Mezincescu, B.E.W. Nilsson and P.K. Townsend,  Phys. Lett. 
 B162 (1985) 116.


\bi{GA}
S. Bellucci, S.J. Gates, Jr., B. Radak, P. Majumdar
   and Sh. Vashakidze,  Mod. Phys. Lett.   A4 (1989) 1985;
S.J. Gates, Jr., P. Majumdar, B. Radak and  Sh. Vashakidze,
    Phys. Lett.    B226 (1989) 237;  B. Radak and Sh. Vashakidze,   Phys. Lett.  B255 (1991) 528.



\bi{RT}
J. Russo and A.A. Tseytlin,
``Green-Schwarz superstring action in a curved magnetic Ramond-Ramond background", JHEP 04 (1998) 014, 
hep-th/9804076.



\bibitem{Ber1}
E. Bergshoeff, E. Sezgin and P.K. Townsend,
Ann. Phys. 185 (1988) 330.



\bi{HM}
M. Henneaux and L.  Mezincescu, ``A $\sigma$-model
interpretation
of Green-Schwarz covariant superstring action",
Phys. Lett. B152 (1985) 340.





\bibitem{N2}
P. van Nieuwenhuizen,
``General Theory of Coset Manifolds and Antisymmetric Tensors
Applied to Kaluza-Klein Supergravity",
in: Proceedings of the
Trieste Spring School on Supersymmetry and Supergravity, Trieste,
   Italy, Apr 4-14, 1984, 
      B. de Wit, P. Fayet, P. van Nieuwenhuizen eds.,
       (World Scientific,
   1984) 
p.239.


\bibitem{N3}
F.A. Bais, H. Nicolai and P. van Nieuwenhuizen,
Nucl. Phys. B228 (1983) 333.

\bibitem{Si1}
W. Siegel, Phys. Rev. D50 (1994) 2799

\bibitem{Ber2}
E. Bergshoeff  and  E. Sezgin,
Phys. Lett. B354 (1995) 256,
hep-th/9504140.

\bibitem{Ber3}
E. Sezgin,
``Super p-Form Charges and a Reformulation of the supermembrane
Action in Eleven Dimensions",
hep-th/9512082.

\bi{WW}
E. Witten, Commun. Math. Phys. 92 (1984) 455.

\bi{SS}
W. Siegel,  Phys. Lett. B128 (1983) 397.

\bi{KKK}
R. Kallosh and J. Kumar, ``Supersymmetry enhancement of 
D-p-branes and M-branes", hep-th/9704189.

\bi{deal}
S. Mukhi, Phys. Lett. B162 (1985) 345;
S.P. De Alwis, Phys. Lett. B164 (1985) 67.


\bi{CT} M. Cvetic and A.A. Tseytlin, Phys. Lett. B366 (1996) 95, 
hep-th/9510097;\\
A.A. Tseytlin, Mod. Phys. Lett. A11 (1996) 689, 
hep-th/9601177


\bi{BG}
T. Banks and M.B. Green,
``Non-perturbative Effects in $AdS_5\times S^5$ String Theory and d=4 SUSY Yang-Mills", 
hep-th/9804170.

\bi{KR}
 R. Kallosh  and  A. Rajaraman, 
``Vacua of M-theory and string theory", 
hep-th/9805041.

\bi{gris}
M.T. Grisaru and D. Zanon, 
Nucl. Phys. B310 (1988) 57;\\
M.T. Grisaru, H. Nishino and D. Zanon, 
Nucl. Phys. B314 (1989) 363.


\bibitem{BRFR}
P. Breitenlohner and D.Z. Freedman
Ann. Phys. 144 (1982) 249.


\bibitem{FGib}
C.J.C.~Burges, D.Z.~Freedman, S.~Davis and G.W.~Gibbons,
Ann. Phys. 167 (1986) 285.


\end{thebibliography}
\end{document}